\documentclass[a4paper,11pt]{article}

\usepackage{graphicx}
\usepackage{xcolor}
\usepackage{todonotes}
\usepackage{authblk}
\usepackage{amsmath}
\usepackage{amsfonts}
\usepackage{subcaption}

\begin{document}

\title{Casimir preserving spectrum of two-dimensional turbulence}

\author[1,2]{Paolo Cifani}
\author[3]{Milo Viviani}
\author[1]{Erwin Luesink}
\author[4]{Klas Modin}
\author[1,5]{Bernard J. Geurts}

 \affil[1]{Multiscale Modeling and Simulation, Faculty EEMCS, University of Twente, P.O. Box 217, 7500 AE Enschede, The Netherlands}
 \affil[2]{Gran Sasso Science Institute, Viale F. Crispi, 7 67100 L’Aquila, Italy}
 \affil[3]{Scuola Normale Superiore di Pisa, Pisa.}%
\affil[4]{Department of Mathematical Sciences, Chalmers University of Technology and University of Gothenburg, 412 96 Gothenburg, Sweden.}%
\affil[5]{Multiscale Energy Physics, CCER, Faculty Applied Physics, Eindhoven University of Technology, P.O. Box 213, 5600 MB Eindhoven, The Netherlands}

\date{\today}

\maketitle

\begin{abstract}
We present predictions of the energy spectrum of forced two-dimensional turbulence obtained by employing a structure-preserving integrator. In particular, we construct a finite-mode approximation of the Navier-Stokes equations on the unit sphere, which, in the limit of vanishing viscosity, preserves the Lie-Poisson structure. As a result, integrated powers of vorticity are conserved in the inviscid limit. We obtain robust evidence for the existence of the double energy cascade, including the formation of the $-3$ scaling of the inertial range of the direct cascade. We show that this can be achieved at modest resolutions compared to those required by traditional numerical methods. 
\end{abstract}

Navier-Stokes equations in two dimensions constitutes a fundamental model for numerous physical phenomena, in particular for large-scale dynamics of geophysical flows. In the limit of vanishing viscosity, two-dimensional fluid dynamics is characterised by an infinite number of first integrals, i.e., the integrated powers of vorticity. This set of constraints, absent in three dimensions, has profound effects on the energy transfer mechanisms across scales of motion \cite{abramov2003statistically}. About half a century ago, Kraichnan \cite{Kraichnan1967} conjectured the coexistence of two inertial ranges for a fluid stirred by a force confined to a typical wavenumber $\kappa_f$. In his theoretical argument, an inverse energy cascade with spectrum $E(\kappa) \simeq \epsilon^{2/3} \kappa^{-5/3}$ for $\kappa \ll \kappa_f$ and a direct energy cascade with spectrum $E(\kappa) \simeq \eta^{2/3} \kappa^{-3}$ for $\kappa \gg \kappa_f$ would be established. Here, $\epsilon$ and $\eta$ are the energy and enstrophy transfer rate, respectively. Both experimental and numerical studies have been conducted to confirm Kraichnan's theory; a comprehensive review is given in \cite{Boffetta2012}. However, despite the ever-increasing computational power and the mounting evidence that points toward the double-cascade scenario, a clear match between theory and numerical results has not yet been established. Particularly cumbersome to capture is the $\kappa^{-3}$ scaling of the direct cascade. In \cite{Boffetta2010} a pseudo-spectral method was applied at extreme numerical resolutions in the quest of verifying Kraichnan's prediction for high Reynolds numbers. While convergence to the $-3$ exponent is indicated, there is still a significant gap with theory even at the highest resolutions included. 

In this work we will show that Kraichnan's scaling of forced two-dimensional turbulence can be accurately captured, at comparably modest numerical resolutions, by employing a geometric numerical method that preserves all invariants of motion in the inviscid limit. In fact, the basis for the theoretical argument that leads to the scaling of the inertial ranges is built upon inviscid conservation laws. It is therefore natural to embed these fundamental properties of the continuum into the numerical algorithm for the discrete system used for the simulation. Based on the seminal work of Zeitlin \cite{Zeitlin1991,Zeitlin2004} and by recent advances \cite{Modin2020,Modin2020_sphere} we construct a finite-mode approximation of the two-dimensional Navier-Stokes equations, which in the limit of vanishing viscosity conserves all discrete integrals of motion, called Casimir functions. % in the language of differential geometry. 

Most numerical studies of homogeneous turbulence were conducted on a periodic square \cite{Lilly1969,Borue1993,Gotoh1998,Lindborg2000,Pasquero2002,Boffetta2007}. Here, motivated by the spectral analysis on a spherical domain \cite{Lindborg2022}, we instead simulate turbulence on the unit sphere. As shown in \cite{Lindborg2022}, the scaling laws of turbulence, traditionally derived on the flat torus, carry over to the sphere where the degree of the spherical harmonic functions, $l$, takes the place of the wavenumber. Furthermore, no artificial boundary conditions have to be imposed, as is the case for 
% simulations on 
the flat periodic domain. The spherical geometry constitutes, then, an ideal numerical testing ground. 

The geometric description of fluid dynamics arises when characterising the motion of inviscid fluid parcels, on manifolds, in the Eulerian viewpoint. We, thus, begin by considering the incompressible Euler equations. Viscous dissipation and forcing will be added later to arrive at the Navier-Stokes equations.
% with which all simulations in this study will be executed. 
The idea is to employ a structure-preseving integrator for convection, which guarantees discrete conservation of Casimirs in the inviscid limit. As a result, balances between convective transport and physical dissipation are not altered by inaccurate discretisation of advection \cite{Verstappen2003}. These balances are, in fact, of fundamental importance for turbulence dynamics. 

Euler's equations of an ideal fluid expressed in the velocity field $v$ are
\begin{equation}
\begin{cases}
\dot{v} + v\cdot\nabla v = -\nabla p, \\
\nabla\cdot v = 0,
\end{cases}
\label{eq:Euler_vel}
\end{equation}
were $p$ is the scalar pressure field. The equations \eqref{eq:Euler_vel} on the 2-sphere $\mathbb{S}^2$ can be written in terms of the vorticity field $\omega(x):=(\nabla\times v(x))\cdot x$, for $x\in\mathbb{S}^2$.
Expressed in the vorticity $\omega$, Euler's equations are
\begin{equation}
\begin{cases}
\dot{\omega} = \{ \psi,\omega \}, \\
\Delta \psi = \omega,
\end{cases}
\label{eq:Euler_vort}
\end{equation}
where $\{\cdot,\cdot\}$ is the Poisson bracket
\begin{equation}
\begin{aligned}
\{ \psi,\omega \}(x)=x\cdot(\nabla\psi\times\nabla\omega), && \forall x\in\mathbb{S}^2,
\end{aligned}
\end{equation}
$\psi$ is the stream function and $\Delta$ is the Laplace-Beltrami operator. 
Euler's equations constitute a Lie--Poisson system on $C^{\infty} (\mathbb{S}^2)$ (see \cite{Arnold_book,Marsden_book}), for the Lie--Poisson bracket given by
\[
\lbrace \mathcal{F}, \mathcal{G}\rbrace_{LP}(\omega)=\int\omega \lbrace \frac{\delta\mathcal{F}}{\delta\omega},\frac{\delta\mathcal{G}}{\delta\omega} \rbrace,\]
$\forall \mathcal{F,G} : C^{\infty} (\mathbb{S}^2)\rightarrow \mathbb{R}$ smooth and Hamiltonian 
\[
\mathcal{H}(\omega)=-\dfrac{1}{2}\int \omega\psi.
\]
Hence, equation \eqref{eq:Euler_vort} is equivalent to
\begin{equation}\label{eq:Euler_vort_LP}
\dfrac{d\mathcal{F}}{dt}(\omega)=\lbrace \mathcal{F}, \mathcal{H}\rbrace_{LP}(\omega),
\end{equation}
$\forall \mathcal{F} : C^{\infty} (\mathbb{S}^2)\rightarrow \mathbb{R}$.
For system (\ref{eq:Euler_vort_LP}) there exists infinitely many Casimir functions $\mathcal{C}: C^{\infty} (\mathbb{S}^2)\rightarrow \mathbb{R}$
\begin{equation}
    \begin{aligned}    
    \{ \mathcal{C},\mathcal{G} \}_{LP} \equiv 0 && \forall \mathcal{G} : C^{\infty} (\mathbb{S}^2)\rightarrow \mathbb{R}.
    \label{eq:Casimir_bracket}
    \end{aligned}
\end{equation}
In particular, the integrated powers of vorticity
\begin{equation}
\begin{aligned}
\mathcal{C}_k(\omega) = \int \omega^k, && k=1,2,\ldots
\end{aligned}
\label{eq:Casimirs}
\end{equation}
are invariants of motion.
In order to derive a discretisation that captures (\ref{eq:Casimir_bracket}) in a discrete sense, it is essential to look at Euler's equations from the geometric viewpoint and embed the underlying differential structure into the discrete system. This process of consistent discretisation is often referred to as \textit{quantisation}, which we will briefly summarise in the following for completeness. 

The dynamics of incompressible ideal fluids can be understood as an evolution equation on the cotangent bundle of the infinite-dimensional Lie group of volume-preserving diffeomorphisms \cite{Arnold_book}. As mentioned, the vorticity equation (\ref{eq:Euler_vort}) constitutes an infinite-dimensional Lie-Poisson system on the space of smooth functions ${C}^{\infty}(\mathbb{S}^2)$. To numerically preserve the Poisson structure we follow the approach of Zeitlin \cite{Zeitlin2004}, which is based on the theory of geometric quantisation pioneered by Hoppe \cite{Hoppe1989} and later analysed in \cite{Bordemann1991,Bordemann1994}. For the purpose here, quantisation refers to the process of constructing a Lie algebra of $N\times N$ complex matrices so that the Poisson bracket is approximated by the matrix commutator. In \cite{Bordemann1994} it was shown that there exists a basis $T_{lm}^N$ of $\mathfrak{u}(N)$ (skew-Hermitian matrices) with structure constants converging to those of the spherical harmonics basis of ${C}^{\infty}(\mathbb{S}^2)$. A projection $\Pi_N : {C}^{\infty}(\mathbb{S}^2) \to \mathfrak{u}(N)$ can be constructed such that, given the basis of standard spherical harmonics $Y_{lm}$ and the basis provided by the matrices $T^{N}_{lm} \in \mathfrak{u}(N)$, for $l=1,...,N$, we have that for $N \to \infty$
\begin{enumerate}
\item $\Pi_N f - \Pi_Ng \to 0$ implies $f=g$,
\item $\Pi_N \{ f,g \} = \frac{N^{3/2}}{\sqrt{16\pi}}[\Pi_Nf, \Pi_Ng] + \mathcal{O}(1/N^2)$,
\item $\Pi_N Y_{lm} = T^{N}_{lm}$,
\end{enumerate}
where $[\cdot,\cdot]$ is the matrix commutator. 
The setting naturally restricts to the subalgebra $\mathfrak{su}(N)\subset \mathfrak{u}(N)$ of traceless matrices, which correspond to zero mean vorticity and stream function.
We thus have a finite-dimensional Lie algebra which converges to that of divergence free vector fields in the sense explained above. 
Using the discrete basis $T^{N}_{lm}$ one can then rewrite (\ref{eq:Euler_vort}) in discrete form \cite{Zeitlin2004,Modin2020_sphere}:
\begin{equation}
\begin{cases}
\dot{W} = [P,W]\\
\Delta_N P = W,
\end{cases}
\label{eq:Euler_vort_mat}
\end{equation}
where $W  \in \mathfrak{su}(N)$ is the vorticity matrix, $P  \in \mathfrak{su}(N)$ is the stream matrix and $\Delta_N$ is the discrete Laplacian given by \cite{Hoppe1998}. Furthermore, as pointed out by Zeitlin \cite{Zeitlin1991}, traces of powers of $W$,
\begin{equation}
\begin{aligned}
C_k(W)=\text{Tr}(W^k) && \text{for } k=1,\ldots,N,
\end{aligned}
\label{eq:power_vort}
\end{equation}
are conserved by system (\ref{eq:Euler_vort_mat}). This is the discrete analogue of conservation of integrated powers of vorticity in the continuum. Moreover, skew-symmetry implies conservation of the total energy. 

To conserve the first integrals of the quantised system (\ref{eq:Euler_vort_mat}) one needs a time discretisation that preserves the Lie-Poisson structure. The key observation is that when the evolution of a skew-symmetric matrix, $W$, can be written in terms of the commutator with another skew-symmetric matrix, then the flow is isospectral, i.e., the eigenvalues of $W$ are constants of motion (cf.~\cite{Hairer_book}). In turn, conservation of the spectrum of $W$ is equivalent to conservation of the Casimirs. Modin and Viviani \cite{Modin2020} recently derived a class of Lie-Poisson integrators for isospectral flows. Their method is based on a discrete Lie-Poisson reduction of symplectic Runge-Kutta schemes, where time-stepping is performed directly at the level of the matrix Lie algebra. This avoids entirely the use of the exponential map, known as the Achille heel of geometric integrators due to its high computational cost. As a result, geometric integration is made viable also for systems with large degrees of freedom, such as fluid systems. As in \cite{Modin2020,Viviani2020}, here we employ the isospectral midpoint method as geometric time integrator.For a detailed description of the numerical method we refer to \cite{cifani2022efficient}. 

We now move on to the Navier-Stokes equations in 2D, which on the unit sphere are written as 
\begin{equation}
\dot{\omega} = \{ \psi,\omega\} + \nu \left( \Delta \omega + 2\omega \right) -\alpha \omega + f,
\label{eq:NS_sphere}
\end{equation}
where $\nu$ is the molecular viscosity, $\alpha \omega$ the damping to avoid accumulation of energy at large scales, $f$ an external forcing and $\nu \omega$ an additional term that arises from the spherical geometry \cite{Lindborg2022}.
Equations \eqref{eq:NS_sphere} can be directly derived from the Navier--Stokes equations in terms of the velocity field $v$, by applying the curl operator $\nabla\times\cdot$.
The forcing is localised in spherical harmonic space in a narrow band around harmonics of degree $l_f$. Moreover, $f$ is assumed to vary over time as white noise so that it is uncorrelated from time-scales of the turbulent flow. This is a common choice for forced homogeneous turbulence in 2D \cite{Boffetta2007} and in 3D \cite{Alvelius1999}. Viscous dissipation and damping are integrated in time by a standard Crank-Nicholson scheme \cite{Crank1947}. By denoting with $\varphi_{\it iso}$ the isospectral map for convection and by $\varphi_{\it CN}$ the Crank-Nicholson map for the remaining terms in (\ref{eq:NS_sphere}), the integration of the vorticity matrix is obtained by means of the second-order Strang splitting \cite{Strang1968}:
\begin{equation}
W^{n+1} = \left(\varphi_{{\it CN},h/2} \circ \varphi_{{\it iso},h} \circ \varphi_{{\it CN},h/2}\right)( W^n),
\label{eq:new_vort_mat}
\end{equation}
with $h$ the time-step and $n$ the time level. Operator splitting is a simple and effective way to keep the geometric structure of convection unaltered by dissipation and forcing. Simulation of the turbulent flow is carried out on a dedicated parallelized Fortran code \cite{cifani2022efficient}. The code makes efficient use of distributed and shared memory by combining MPI \cite{Gropp1996} with multithreading. 

\begin{figure}[hbt!]
\centering
\includegraphics[width=.47\textwidth]{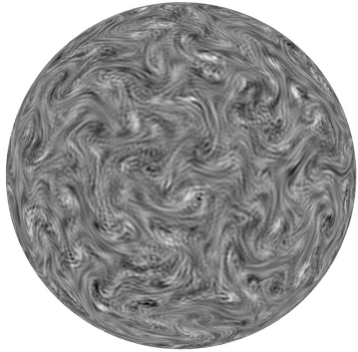}
\caption{Vorticity field at statistically steady state. Vorticity values increase from light to dark gray.}
\label{fig:vort_field}
\end{figure}

A qualitative illustration of the flow is provided in Fig.~\ref{fig:vort_field} by a snapshot of the vorticity field at statistically steady state.

\begin{figure}[hbt!]
\centering
\includegraphics[width=.45\textwidth]{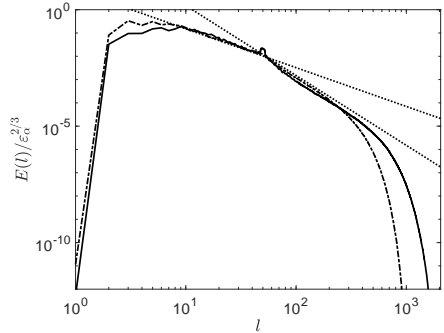}
\caption{Kinetic energy spectrum, scaled by the friction energy dissipation rate $\varepsilon_\alpha$, as a function of $l$ for $N=1024$ (dash-dotted line) and $N=2048$ (solid line). The Kraichnan scalings $-5/3$ and $-3$ are shown as reference.}
\label{fig:E_sp}
\end{figure}

In Fig.~\ref{fig:E_sp} we present the spectrum of turbulence, scaled by energy dissipation rate $\varepsilon_\alpha$ associated with the friction term in (\ref{eq:NS_sphere}), at resolutions $N=1024$ and $N=2048$. Analogously to \cite{Boffetta2010}, the forcing is placed at spherical harmonic index $l_f=50$ to enable the development of the two inertial ranges over about a decade toward large scales ($l \ll l_f$) as well as toward small scales ($l \gg l_f$). The damping coefficient $\alpha$ is tuned to balance the fraction of the energy injected by the forcing and transferred at large scales so that a statistically stationary state is reached. In both simulations, resolution is set to have $NL_\nu \approx 1$, with $L_\nu = \nu^{1/2} / \eta_\nu^{1/6}$ the enstrophy dissipation scale \cite{Boffetta2007}. This corresponds to a proper resolution of the flow, which can be appreciated by the decay of the spectrum at $l \approx N$ where the enstrophy content goes down to machine precision. The chosen setting enables us to simulate two Reynolds numbers for the direct cascade \cite{Kraichnan1967}, Re$=\left[ E(l_f)/l_f \right]^{1/2}/\nu$, equal to $1550$ for $N=1024$ and $5200$ for $N=2048$.

\begin{figure}[hbt!]
\centering
\includegraphics[width=.45\textwidth]{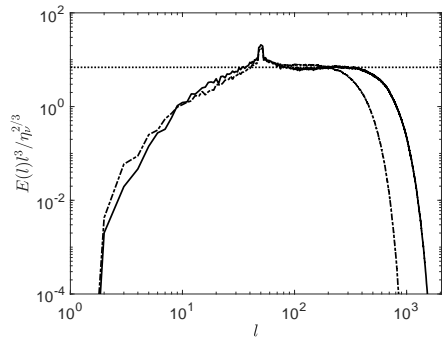}
\caption{Kinetic energy spectrum, scaleded by the enstrophy dissipation rate $\eta_\nu$ and compensated by $l^{3}$, as a function of $l$ for $N=1024$ (dash-dotted line) and $N=2048$ (solid line). The computed value of $C'=6.8$ is shown by the horizontal line.}
\label{fig:E_sp_comp}
\end{figure}

The $-5/3$ inertial range is directly visible in the spectrum for $l<l_f$, up to the largest scales of the flow where damping is dominant. The computed value of the Kolmogorov constant $C$, of the scaling law $E(l)=C\epsilon^{2/3}_\alpha l^{-5/3}$, is about $8.2$. This is in line with the estimates reported in literature obtained from simulations on the flat torus \cite{Boffetta2000,Boffetta2010}, with values ranging from $6$ to $7$. More interestingly, our results show a robust $-3$ scaling for $l>l_f$. This is clearly visible from the compensated spectrum of Fig.~\ref{fig:E_sp_comp}. The estimated value of the constant $C'$ in the Kraichnan scaling law $E(l)=C'\eta^{2/3}_\nu l^{-3}$ is approximately $6.8$ and found quite independent on the resolutions here investigated. In fact, the computed least-square estimate of the exponent in the inertial range of the direct cascade is $-3.10$ for $N=1024$ and $-3.02$ for $N=2048$, suggesting converge toward the theoretical value for increasing Re. Evidence of the $-3$ scaling was also found in numerical studies of Earth's atmosphere \cite{augier2013new}. As for the additional logarithmic correction postulated by Kraichnan \cite{Kraichnan1971} to account for non-local energy transfer, our results suggest that these non-local transfers have a negligible effect on the actual turbulence spectrum. 

\begin{figure}[hbt!]
\centering
\begin{subfigure}[b]{0.4\textwidth}
\includegraphics[width=1\textwidth]{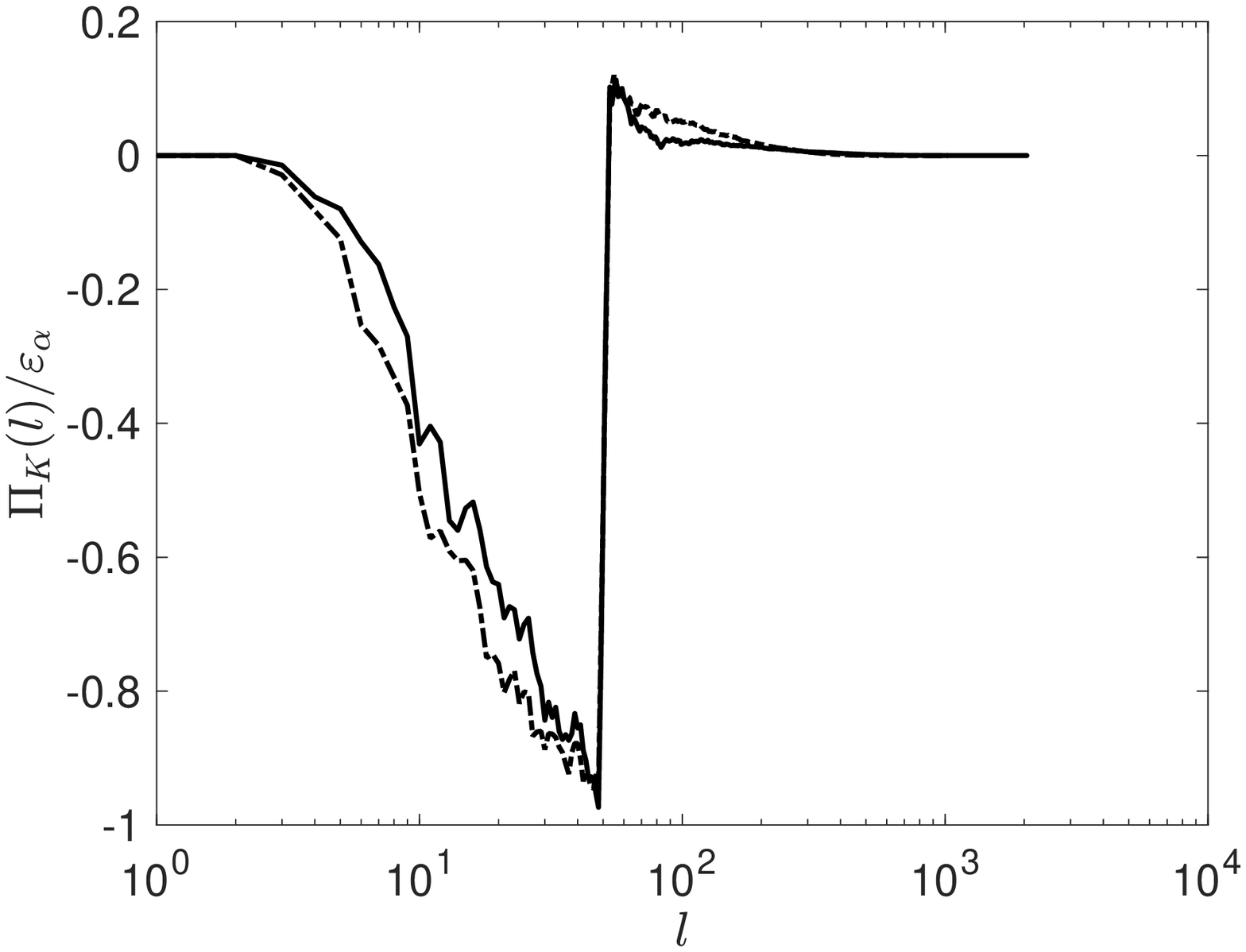}
\end{subfigure}
\hfill
\begin{subfigure}[b]{0.4\textwidth}
\includegraphics[width=\textwidth]{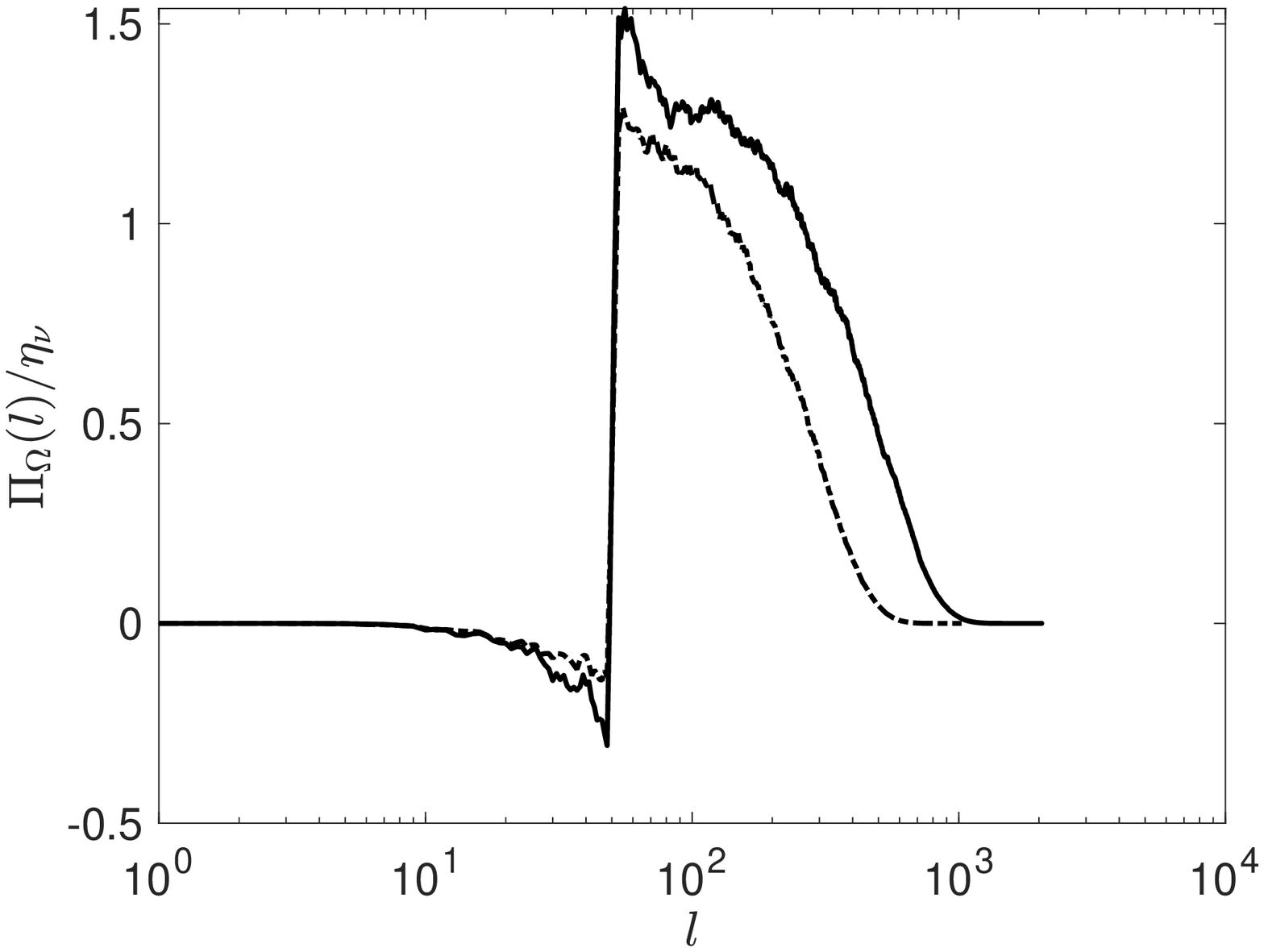}
\end{subfigure}
\caption{Spectral convective energy fluxes (top panel) and enstrophy fluxes (bottom panel) as a function of $l$ for $N=1024$ (dash-dotted lines) and for $N=2048$ (solid lines).}
\label{fig:fluxes}
\end{figure}

The existence of a double cascade can be further inferred from the spectral convective energy flux $\Pi_K(l)$ and enstrophy flux $\Pi_\Omega(l)$, defined as in \cite{Boffetta2010} and presented in Fig. \ref{fig:fluxes}. Net energy transfer is observed for $l<l_f$ with $\Pi_K(l) \approx 0$ for $l \gg l_f$ while net enstrophy transfer takes place for $l>l_f$ with $\Pi_\Omega(l) \approx 0$ for $l \ll l_f$. The chosen forcing in spectral space causes a sharp transition around $l_f$. For the high resolution case, the formation of an approximately constant enstrophy flux $\eta_\nu$ can be observed, as predicted by Kraichnan's theory. As shown in \cite{Boffetta2010}, we expect the $l$-range in which $\Pi_\Omega(l) \approx \eta_\nu$ to extend further for higher resolutions as the dissipative scale is separated further from the scale of the forcing. Analogous considerations can be made for the energy fluxes $\Pi_K(l)$.

In conclusion, in this letter we provided robust evidence for the existence of the double cascade in forced two-dimensional turbulence after half a century its theoretical development. We simulated the Navier-Stokes equations on a sphere explicitly preserving the differential geometric structure of convection in the limit of vanishing viscosity. Contrary to traditional numerical methods, where the focus is on minimising the local truncation error, here we followed mathematically motivated principles of embedding conservation of first integrals into the numerical scheme. Predictions of the double cascade of the energy spectrum were confirmed at modest resolutions compared to the pioneering work in \cite{Boffetta2010}, where resolutions at $N=32768$ were found necessary to achieve first indications of the double scaling, both on the scaling and regarding the virtues of geometric methods for fluids. Conservation of enstrophy is arguably crucial for the correct prediction of the direct cascade. Whether conservation of higher powers of vorticity leads to more accurate computations of the spectrum of turbulence or of higher order statistics remains an interesting subject of future research.

%%%%%%%%%%%%%%%%%%%%%%%%%%%%%%%%%%%%%%

\section*{Acknowledgements}
This publication is part of the project SPRESTO which is financed by the Dutch Research Council (NWO). 
This work was also supported by the Swedish Research Council, grant number 2017-05040, and the Knut and Alice Wallenberg Foundation, grant number WAF2019.0201.
Simulations were carried out on the Dutch national e-infrastructure with the support of SURF Cooperative. 
We wish to thank prof.\ Guido Boffetta for his valuable suggestions in the setup of the simulations. 

\bibliographystyle{plain}
\bibliography{biblio}

\end{document}